\documentclass[onecolumn]{aa520}
\usepackage{graphicx}

\begin{document}
\headnote{Research Note}
   \title{A direct and differential imaging search for sub-stellar companions to $\epsilon$ Indi A 
\thanks{Based on observations made with ESO Telescopes at the Paranal Observatories under programme ID 273.C-5030 and on observations made with the NASA/ESA Hubble Space Telescope, obtained from the data archive at the Space Telescope Institute. STScI is operated by the association of Universities for Research in Astronomy, Inc. under the NASA contract NAS 5-26555.}}

   \author{K. Gei{\ss}ler\inst{1,2} \and
   	  S. Kellner\inst{1,3} \and
   	  W. Brandner\inst{1} \and
          E. Masciadri\inst{1,4} \and
          M. Hartung\inst{2} \and
          T. Henning\inst{1} \and
          R. Lenzen\inst{1} \and
          L. Close\inst{5} \and
          M. Endl\inst{6} \and
          M. K\"urster\inst{1}
          }


   \institute{Max-Planck-Institut f\"ur Astronomie, K\"onigstuhl 17,
              D-69117 Heidelberg, Germany\\
              \email{kgeissle@eso.org, skellner@keck.hawaii.edu, brandner@mpia.de}
         \and European Southern Observatory, Alonso de Cordova 3107, Vitacura, Santiago, Chile
	 \and W.M. Keck Observatory, 65-1120 Mamalahoa Hwy, Kamuela, Hawaii  96743
	 \and INAF - Osservatorio Astrofisico di Arecetri, L.go E. Fermi 5, 50125 Firenze, Italy
         \and Steward Observatory, University of Arizona, Tucson, Arizona 85721, USA
         \and McDonald Observatory, The University of Texas at Austin, Austin, TX 78712-1083, USA\\         
             }

   \date{Received <date>; accepted <date>}

   \abstract{
We have carried out a direct and differential imaging search for sub-stellar companions to $\epsilon$\,Indi\,A using the adaptive optics system NACO at the ESO VLT. The observations were carried out in September 2004 with NACO/SDI as well as with NACO's S27 camera in the H and $K_{s}$ filters. The SDI data cover an area of $\sim$\,2.8\arcsec around $\epsilon$ Indi A. No detection was achieved in the inner neighbourhood down to 53\,$\rm M_{J}$ (5\,$\sigma$ confidence level) at a separation $\geq$\,0.4\arcsec(1.45\,AU) and down to 21\,$\rm M_{J}$ for separations $\ge$ 1.3" (4.7 AU). To cover a wider field of view, observations with the S27 camera and a coronagraphic mask were obtained. We detected a faint source at a separation of (7.3\,$\pm$\,0.1)\arcsec and a position angle of $(302.9\,\pm\,0.8)^{\circ}$. The photometry for the candidate companion yields $\rm m_{H}$\,=\,(16.45\,$\pm$\,0.04)\,mag and $\rm m_{K_{s}}$\,=\,(15.41\,$\pm$\,0.06)\,mag, respectively. Those magnitudes and the resulting color (H-$\rm K_{s}$) = (1.04\,$\pm$\,0.07)\,mag fit best to a spectral type of L5\,-\,L9.5 if it is bound.\newline Observations done with HST/NICMOS by M. Endl have shown the source to be a background object.  
   \keywords{Stars: brown dwarfs --
             Stars: fundamental parameters
               }
   }

   \maketitle
%

\section{Introduction}

At present, the nearby, high proper motion star $\epsilon$\,Ind\,A has been a focus for studies about brown dwarfs and extrasolar planets for two different reasons. Firstly, since 2003 (Scholz et al., 2003) $\epsilon$ Ind A has been known to have a physical companion at a separation of $\sim$\,1500\,AU which thereafter was found to be a binary consisting of two T dwarfs in the following named $\epsilon$\,Ind Ba and Bb (Volk et al.$\ \cite{volk03}$, McCaughrean et al.$\ \cite{mccaughrean04}$). The two components have spectral types of T1 and T6 and estimated masses of 47 and 28\,$\rm M_{Jup}$, respectively, assuming an age of 1.3\,Gyr for the system. Dominated by the uncertainty in the age determination (possible range 0.8\,-\,2\,Gyr; Lachaume et al.$\ \cite{lachaume99}$), the uncertainties in the masses were calculated to be $\le$\,25\,$\%$. Secondly, Endl et al. (2002) reported the detection of a linear trend in the radial velocity measurements of $\epsilon$\,Ind\,A. The observations were carried out with the ESO Coud\'e Echelle Spectrometer (CES) on La Silla. The best-fit to the RV measurements gives an RV shift of +\,0.012\,$\pm$\,0.002\,m$\rm s^{-1}$ per day. The rms scatter around the slope is given as 11.6\,m$\rm s^{-1}$. Endl et al. ($\cite{endl02}$) considered that the linearity of the trend could be caused by a distant stellar companion as well as by a very long-period (P\,$>$\,20\,yrs) planetary companion. Our observations aimed at the direct detection of this companion.
\begin{table*}[htb]
\caption[]{Parameters of $\epsilon$\,Ind\,A}
\label{star}
\begin{tabular}{l r}
\hline
\noalign{\smallskip}
\hline           
distance &   (3.626\,$\pm$\,0.009)\,pc \\
spectral type & K4.5\,V \\
proper motion & $\sim\,4.7$\,as/yr \\
age & $\sim\,1.3$\,Gyr \\
age range & (0.8\,-\,2)\,Gyr \\
mass & 0.7\,$\rm M_{\odot}$ \\
H magnitude & 2.35\,mag \\
K magnitude & 2.24\,mag \\
\noalign{\smallskip}
\hline
\end{tabular}
\end{table*}
\section{Observations and data reduction}

\subsection{Observation}

In September 2004 observations of a sample of young, nearby stars were carried out at the VLT UT4 (Yepun) using the SDI (Simultaneous Differential Imager) mode of NACO. SDI has been implemented into CONICA (Lenzen et al.$\ \cite{lenzen03}$) and has been commissioned in August 2003 and February 2004. The special design of SDI allows simultaneous observations of four images through three narrow band filters, which are centered around the methane absorption band at 1.6 microns\protect\footnote{The three narrow band filters are located at 1.575$\mu$m , 1.60$\mu$m and 1.625$\mu$m.} (Lenzen et al.$\ \cite{lenzen04}$). 
The image scale of the SDI camera is 17.25 mas/pixel and the total field-of-view (FOV) is $5\,\arcsec\,\times\,5\,\arcsec$, but due to the misalignment of the SDI field mask it  was $2.7\,\arcsec\,\times\,3.7\,\arcsec$ at the time of our observations. For each star two SDI data sets were obtained at two rotator angles of $0^{\circ}$ and $33^{\circ}$. $\epsilon$\,Ind\,A was observed on September, 18. Per rotator angle 15 images were taken, each frame with a total integration time of 0.5\,sec\,$\times$\,192\,=\,96\,sec, resulting in a combined integration time of 24\,min per observation angle.
 During the observations of $\epsilon$\,Ind\,A the mean seeing was 1\arcsec and the average Strehl ratio was 31$\%$.
\newline
Due to the limited FOV and due to the knowledge of the radial velocity trend reported by Endl et. al. ($\cite{endl02}$) indicating a potential companion at large distances, additionally images with the NACO S27 objective were taken on September, 19. to cover a wider FOV. This objective has a FOV of $28\,\arcsec\,\times\,28\,\arcsec$ with a pixel scale of 27.03 mas/pixel. The observations were done in H and $\rm K_{s}$, with $\epsilon$\,Ind\,A masked out by a $1.4\,\arcsec$ diameter coronographic mask. In each filter two images were obtained, each image with a total integration time of 120\,sec in H and 60\,sec in $\rm K_{s}$, respectively. 
\subsection{SDI data reduction}

\begin{figure}[htb]
\begin{center}
\includegraphics[width=0.6\textwidth,angle=90]{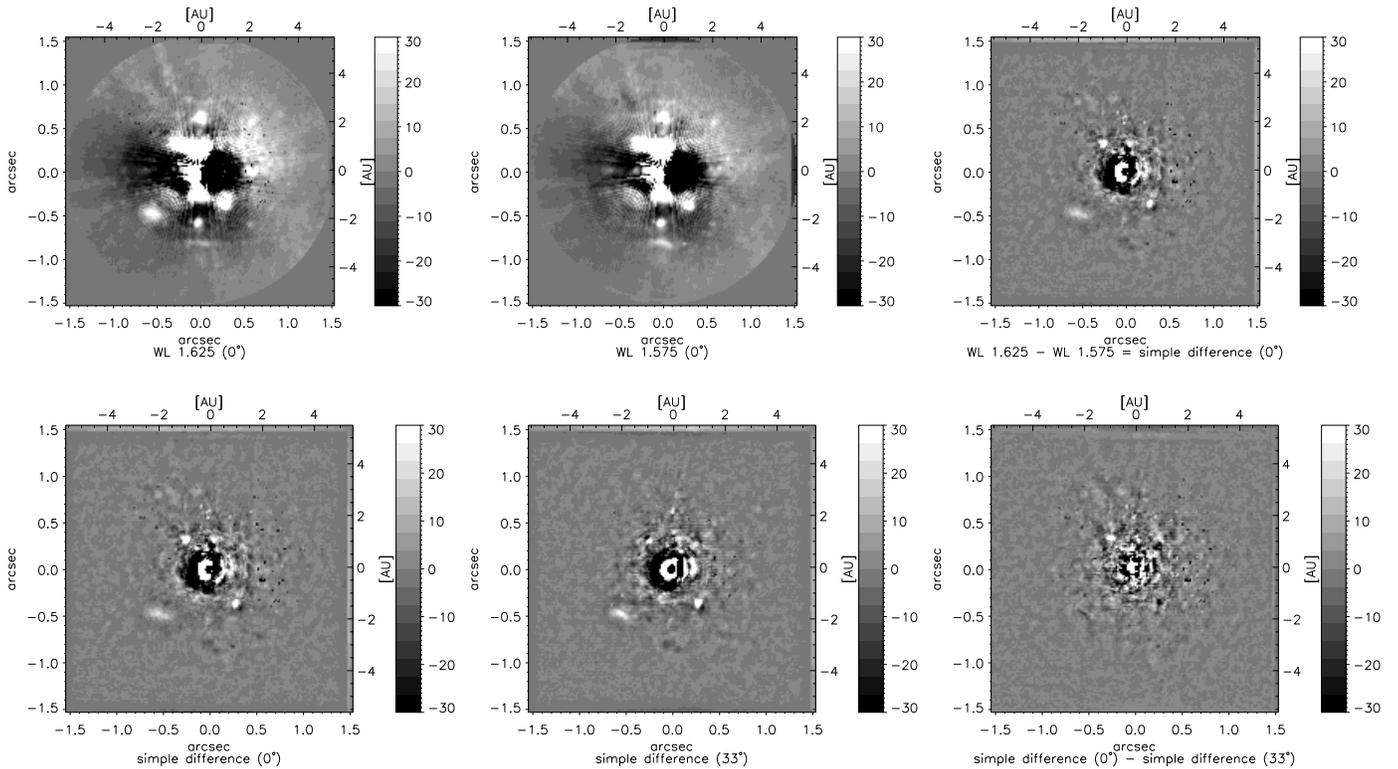}
\end{center}
\caption{SDI reduction sequence. Upper panel: Images obtained in two different narrow band filters at the same rotator angle (here 0$^\circ$) are subtracted to get a "simple difference" (at 0$^\circ$). Note that from the left and middle image the respective symmetric radial profile was subtracted, to improve the visibility of the PSF noise. \newline
Bottom panel: Two "simple difference" frames of different rotator angles (here 0$^\circ$ and 33$^\circ$) are subtracted from each other to increase the contrast and to remove non common path aberrations.}
\label{sdi}
\end{figure}

The SDI data set has been reduced by an inhouse built data reduction pipeline optimized for NACO-SDI purposes. In a first step a standard reduction including flat fielding, sky-subtraction and badpixel mask multiplication is performed with each frame. Since each SDI raw frame contains four images of the star taken simultaneously in three narrow band filters (at 1.575$\mu$m , 1.60$\mu$m and 1.625$\mu$m) the images are cut out and ordered according to their wavelength.\protect\footnote{Two images are taken in the 1.625$\mu$m filter, but one is too close to the edge of the detector and therefore not used.} 
After rescaling them to a common $\lambda / D $ platescale they are flux-calibrated. During the commissioning of the SDI camera, the optical distortion was shown to be less than 0.5\,$\%$ between and inside the given FOVs. Since the effect is very small we did not correct it. In the following all frames which belong to a certain wavelength are added with an accuracy of 0.1\,pixel and averaged. In the next step the exact relative shift between the averaged frames of two different wavelength bands is calculated and thereafter the frames are subtracted to obtain a "simple difference" (see figure $\ref{sdi}$). For each star two SDI data sets with different rotator angles are obtained and separately reduced to obtain a "simple difference" frame for each rotator angles. The "simple difference" frames are in the following used to decrease non common path aberrations, which are still limiting the achieved contrast. While non common path aberrations artefact's are not affected by a rotation of the instrument, objects on sky rotate according to the specified angle. Therefore, they can be eliminated by subtracting two "simple difference" frames at different position angles of the detector. In the first place, to detect possible methane rich companions, we used the "simple difference obtained by subtracting the 1.625$\mu$m from the 1.575$\mu$m narrow band image, since a methane rich companion would be brightest in this difference image.

\subsection{NACO/S27 - wide field data}

\begin{figure}[htb]
\begin{center}
\includegraphics[width=0.3\textwidth,angle=90]{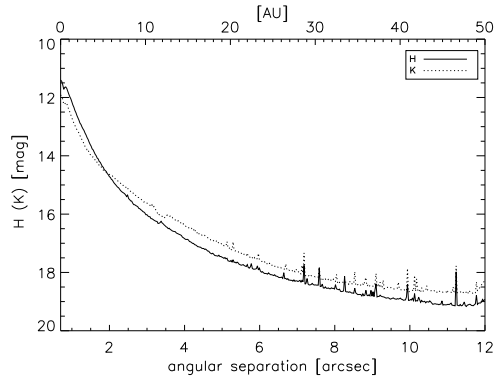}
\end{center}
\caption{5$\sigma$ detection limit for the NACO/S27 observations in H (full line, t\,=\,120\,sec) and $\rm K_{s}$ (dotted line, t\,=\,60\,sec). The region within a radius of 1\arcsec\, from the star should be excluded since $\epsilon$\,Ind\,A was masked out by a $1.4\,\arcsec$ diameter coronographic mask and secondary the area around the coronographic mask is dominated by strong residuals.}
\label{s27_det}
\end{figure}

Common standard reduction steps like sky subtraction, flatfielding and bad pixel correction were applied to the data taken with the S27 camera. To obtain contrast curves for H and $\rm K_{s}$ (see figure $\ref{s27_det}$) the standard deviation in dependence of the radial separation from $\epsilon$\,Ind\,A was calculated.
\section{Analysis}

\subsection{Detection limits for NACO/SDI}

\begin{figure}[htb]
\begin{center}
\includegraphics[width=0.3\textwidth,angle=90]{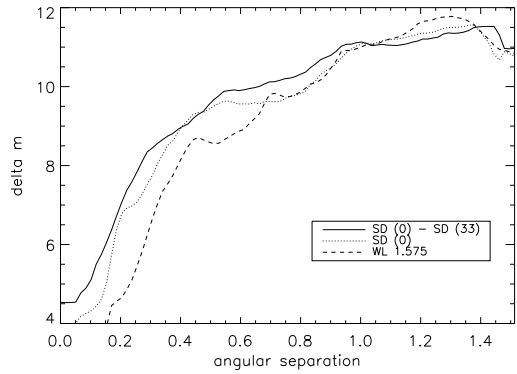}
\end{center}
\caption{The figure shows the 5\,$\sigma$ detection limits for $\epsilon$\,Ind\,A in the narrow band filter at 1.575\,$\mu$m, the "simple difference" (SD) at $0^{\circ}$ and the detection limit after subtracting the two "simple difference" at $0^{\circ}$ and $33^{\circ}$ . In the finally reduced SDI images companions down to 21\,$M_{J}$ (5\,$\sigma$ confidence level, system age of 1\,Gyr) at separations greater than 1.3\arcsec\, should be detectable.}
\label{det}
\end{figure}

The wavelength bands of the SDI camera are centered around the methane absorption band at 1.6 microns, the three narrow band filters are located inside and outside the methane feature (1.575, 1.6, 1.625\,$\mu$m). By subtracting the images of two different wavelength bands the flux of the central star cancels out, meanwhile the flux of a methane rich companion outside the absorption band remains ("simple difference") (Racine et al. $\cite{racine99}$, Marois et al. $\cite{marois00}$, Sparks \& Ford $\cite{sparks02}$) . The two "simple difference" frames (one at $0^{\circ}$ and one at $33^{\circ}$) are subtracted from each other, so that a methane rich companion would appear as a pair of positive and negative spots. In the completely reduced image of $\epsilon$\,Ind\,A such a pair was not detectable.  Figure $\ref{det}$ shows the detection limit obtained in the narrow band filter at 1.575\,$\mu$m as well as in the "simple difference" image. To obtain the contrast curve, the standard deviation of the  residuals was calculated as a function of radial distance from the star.
Before comparing the observed magnitudes to evolutionary models to obtain a mass estimate, the observed detection limits have to be corrected for the offset between the SDI narrow band filters and the H band. This was done by Biller et al. $\cite{biller06}$). The authors used the spectra of 15 objects with spectral types of T\,4.5\,-\,T7 and computed the expected H band magnitude as well as the narrow band magnitudes. From those magnitudes they obtained the offset between the filters for different spectral types. To obtain the mass estimate given in figure $\ref{det}$ we used the offset found for a spectral type of T5, which is (0.5\,$\pm$\,0.05)\,mag and compared the final magnitude to the "COND"\,-\,models by Baraffe et al.$\ \cite{baraffe03}$.

\subsection{Candidate companion}

\begin{figure}[htb]
\begin{center}
\includegraphics[width=8cm,angle=90]{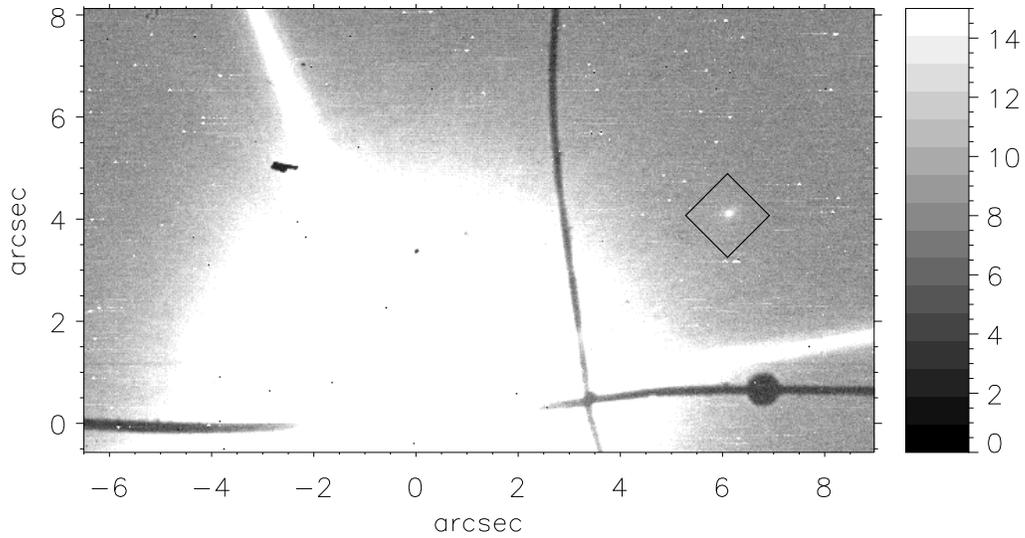}
\end{center}
\caption{Image of $\epsilon$\,Ind\,A taken with NACO's S27 camera in $\rm K_{s}$ and $\epsilon$\,Ind\,A masked-out. A faint source is visible on the upper right corner. It is (7.3\,$\pm$\,0.1)\,$^{\prime\prime}$ away from the center and $(302.9\,\pm\,0.8)^{\circ}$ from the north axis. North is up, East is left.}
\label{freddi}
\end{figure}

\begin{figure}[htb]
\begin{center}
\includegraphics[width=0.5\textwidth,angle=0]{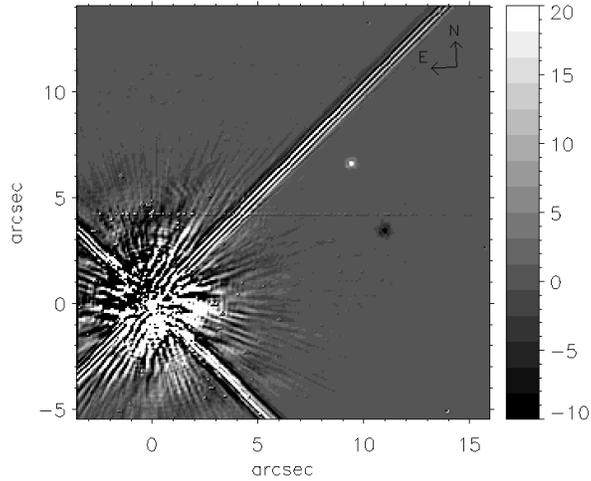}
\end{center}
\caption{HST/NICMOS observations from September,18. 2005. The observations were done in the F160W filter at two different rotator angles (86.8$^\circ$ and 104.8$^\circ$) using the coronographic mode of the nic2 camera. Shown is the by 90$^\circ$ rotated difference image obtained after subtracting the two images of different rotator angles from each other. The former candidate is visible at a separation of (11.3\,$\pm$\,0.2)\arcsec. The north-east orientation is given for the positive source.}
\label{nicmos}
\end{figure}

In the data set obtained with NACO's S27 camera a faint source (see figure $\ref{freddi}$) at a separation of (7.3\,$\pm$\,0.1)\,arcsec ((26.6\,$\pm$\,0.5)\,AU) and a position angle of (302.9\,$\pm\,0.8)^\circ$ was detected, with a S/N ratio greater than 25 in both filters.\protect\footnote{To obtain the separation and the position angle of the source relative to $\epsilon$\,Ind\,A, the pixelscale and N-orientation provided in the image header was used.} The photometry was done with the IRAF APPHOT tool using as reference the standard star GSPC\,S279F. The apparent magnitudes for $\rm K_{S}$ and H are: $\rm m_{K}$\,=\,(15.41\,$\pm$\,0.06)\,mag and $\rm m_{H}$\,=\,(16.45\,$\pm$\,0.04)\,mag, respectively. An approximate spectral type for the candidate companion can be determined by comparing the (H-$\rm K_{s}$) color to the (H-K) color of known brown dwarfs given by Leggett et al. ($\cite{leggett02}$). The color fits best to a spectral type of L5\,-\,L9.\newline
To verify whether the source is a physical companion or a background object we analysed older NICMOS data of $\epsilon$\,Ind\,A taken in August 1997 with the NIC2 camera in the medium band filters F180M, F207M and F222M. In none of the images the source was detected.\newline
New observations done on July, 18. 2005 by M. Endl with HST/NICMOS confute the companionship of $\epsilon$\,Ind\,A and the candidate . The observations were carried out using the coronographic mode of the NIC2 camera and the F160W filter. $\epsilon$\,Ind\,A was observed at two different rotator angles (86.8$^\circ$ and 104.8$^\circ$) with a total integration time of 299.1\,sec at each angle. In both images the source is clearly visible at a separation of (11.3\,$\pm$\,0.2)\arcsec from $\epsilon$\,Ind\,A (see figure $\ref{nicmos}$). In the 10 months between the VLT/NACO observations (Sep. 19, 2004)  and the HST/NICMOS observations (July 18, 2005), $\epsilon$\,Ind\,A moved by about 3.3" to the  east, and 2.1" to the south (i.e. 3.9" in the direction PA\,=\,122.6$^\circ$). Hence the change in separations between the faint source and $\epsilon$ Ind\,A is entirely explained by the proper motion of $\epsilon$\,Ind\,A. The position angle, however, remained almost constant as the direction of proper motion (122.6$^\circ$) is almost perfectly along a line running through the faint source and $\epsilon$\,Ind\,A. The source is definitely not co-moving with $\epsilon$\,Ind\,A.
\begin{table*}[htb]
\caption[]{Parameters of candidate}
\label{planet}
\begin{tabular}{l r r}
\hline
\noalign{\smallskip}
\hline   
 &  NACO/S27 & NICMOS \\
\hline  
date of obs. & 19.09.2004 & 18.07.2005 \\    
separation (\arcsec) & 7.3\,$\pm$\,0.1 & 11.3\,$\pm$\,0.2 \\
position angle $(^\circ)$ & 302.9\,$\pm$\,0.8 & 301.6\,$\pm$\,1.8 \\
H [mag] & 16.45\,$\pm$\,0.04 & - \\
K [mag] & 15.41\,$\pm$\,0.06 & - \\
F160W [mag] & - & 16.93\,$\pm$\,0.05 \\
\noalign{\smallskip}
\hline
\end{tabular}
\end{table*}
\section{Conclusion}

The SDI observations are basically limited by the rather old age of $\epsilon$\,Ind\,A (compared to other SDI target stars), since the luminosity of sub-stellar companions strongly depends on the age of the system. Taking the COND models by Baraffe et al.$\ \cite{baraffe03}$ to estimate the masses, candidates 21\,$\rm M_{J}$ at separations $\geq$\,1.3\,$^{\prime\prime}$(4.7\,AU) should have been discovered. \newline
The observations done with the NACO/S27 camera should have revealed objects with $\rm \,K_{s}\,<\,15.6\,mag$ as well as H\,$<16$\,mag at separations $>$\,3$^{\prime\prime}$ from $\epsilon$\,Ind\,A. When comparing these magnitude limits to the "COND"\,-\,models by Baraffe et al. ($\cite{baraffe03}$) a rough mass limit of (16\,$\pm$\,4)\,$\rm M_{J}$ can be derived.\newline
 The only candidate companion, which could be detected in the NACO/S27 observations, later on was revealed to be a background source by observations done with HST/NICMOS.



\end{document}